# Scattering and Absorption Control in Biocompatible Fibers towards Equalized Photobiomodulation

J. George,[1] H. Haghshenas,[1] D. D'Hemecourt,[1] W. Zhu[2], L. Zhang[2], V. Sorger[1,*]

[1]Department of Electrical and Computer Engineering, The George Washington University, Washington, D.C. 20052, USA
[2]Department of Mechanical and Aerospace Engineering, The George Washington University, Washington, D.C. 20052, USA

*Corresponding author: sorger@gwu.edu

**Abstract:**

**Transparent tissue scaffolds enable illumination of growing tissue to accelerate cell proliferation and improve other cell functions through photobiomodulation. The biphasic dose response of cells exposed to photobiomodulating light dictates that the illumination to be evenly distributed across the scaffold such that the cells are neither under nor over exposed to light. However, equalized illumination has not been sufficiently addressed. Here we analyze and experimentally demonstrate spatially equalizing illumination by three methods, namely, engineered surface scattering, reflection by a gold mirror, and traveling-waves in a ring mesh. Our results show that nearly equalized illumination is achievable by controlling the light scattering-to-loss ratio. This demonstration opens opportunities for dose-optimized photobiomodulation in tissue regeneration.**

**Main Body:**

Photobiomodulation is the use of light to regulate biological activity in cellular or tissular level. Light at red and near-infrared (NIR) frequencies has shown to promote cell proliferation in both *in vitro* and *in vivo* experiments [1, 2]. There are three potential mechanisms for the increased proliferation. The first mechanism is the direct absorption of red or NIR photons by cytochrome c oxidase in the mitochondria of the cells. This absorption increases the rate of the electron transport chain, and consequently the available adenosine triphosphate (ATP), cellular energy, of the cell, increasing the proliferation rate. The second potential mechanism is that red light (670nm and 633nm) decreases the viscous friction of interfacial water layers surrounding the mitochondrial nano-motor, thus speeding up the mitochondrial production of ATP [3]. The third mechanism is that light generates low levels of reactive oxygen species (ROS) within the cell, triggering a survival mechanism that increases the production of mitochondria. This increase in mitochondria then adds to the available ATP within the cell, which increases proliferation.

This third mechanism is optical dose dependent. With too small a dose of light, insufficient ROS will be generated and the cell will not increase its production of mitochondria. On the other hand, if the dose is too high, too much ROS will be generated and the cell will enter apoptosis. As such, the dose response in cell photobiomodulation has been shown to follow a biphasic response curve [4], growing incrementally with increased dosage and leveling at some optimal range before falling precipitously (Inset, Fig 1). This biphasic dose response requires care in the selection of power for *in vivo* use. If tissue is exposed from the outside of the patient, the cells close to the source will receive exponentially more dosage than cells farther from the source, due to Beer's Law of absorption. This optical absorption is not straightforward function of tissue type for all tissues between the source and the target cells. It is therefore reasonable to expect that in a complex organism only a small number of subsurface target cells will reach the exact dose required for optimal effect, while others remain either unaffected or damaged. Addressing such optimized dosage of light delivery to cells deep inside tissue scaffolds and ultimate to the body is the motivation of the work presented here.

Tissue scaffolds are mechanical structures with engineered morphology used to hold cells while they develop to create artificially structured tissues. Tissue scaffolds are created from biocompatible materials such as hydrogels, poly-lactic acid (PLA), and silk fibroin by 3D printing, molding, lithography and other manufacturing techniques [5-8]. Transparent hydrogel scaffolds have been combined with optics for sensing, optogenetics, and photobiomodulation [9]. While the transparent scaffolds do allow light to enter the tissue more easily, they can suffer from the same dose dependence seen in external photobiomodulation where uneven illumination of the growing tissue causes some cells to be overexposed and others to be underexposed.

In this paper we introduce and test three illumination methods for engineering transparent tissue scaffolds with equalized power profiles. These methods can be applied with minimal changes to the mechanical structure of the scaffold, allowing them to be added to existing scaffold designs. The first proposed method utilizes engineered diffusion; optical diffusion is the scattering of light from surface boundaries within a media. The light-loss inside a waveguide depends on both the internal absorption [10-12], scattering events, and any other loss mechanisms such as evanescent coupling [13], bending losses in ring structures, and proximity to metals found in the field of plasmonics [14-18]. Regarding the material loss leading to internal waveguide absorption, this can be in principle complex if the waveguide consists of a multitude of materials where an effective mode index determines the modal loss [19]. Here, however, we only consider waveguides comprised of homogeneous materials. Let us now consider such as homogeneous waveguide comprised entirely of the biocompatible material PLA extending in the x direction into the scaffold. In this case the optical power within the waveguide can be modeled by the differential equation, Eq. 1.

$$\frac{dP}{dx} = (-\alpha - \sigma)P \qquad (1)$$

where light is only lost by absorption, $\alpha$, or by scattering into the fiber surrounding scaffold, $S$. Solving Eq. 1 for the power $P$ at any point within the waveguide is given by Eq. 2.

$$P(x) = Ae^{(-\alpha-\sigma)x} \ [W] \qquad (2)$$

The optical power along the length of the waveguide decays exponentially with distance in the absence of gain. We wish to design the scattering along the length of the waveguide in order to equalize the distribution of power over the surface of the waveguide. Therefore, we seek a scattering profile, $S(x)$, as a function of distance along the x-axis that provides the least power difference between input and end of the fiber, $\min\{P(0) - P(L-\epsilon)\}$, with no power left at the end, $P(L) = 0$, for maximum efficiency. An optimally flat irradiance will therefore have a derivative of zero. If we set the derivative of the irradiance to zero we need to find a function, $S(x)$, that meets the following two criteria:

$$I(x) = \sigma A e^{(-\alpha-\sigma)x} \left[ W \cdot m^{-1} \right] \quad (3)$$

$$\frac{dI}{dx} = \sigma(-\alpha-\sigma) A e^{(-\alpha-\sigma)x} \left[ W \cdot m^{-2} \right] = 0 \quad (4)$$

When we naively attempt to match to the inverse of the exponential function, for example $S(x) = e^{mx}$, we see that in turn increases the exponential decay function. In fact the only function $S(x)$ that can satisfy (3) and (4) for all x is $S(x) = 0$, which violates our requirement that the power at the end of the waveguide reaches 0. Hence the optimum we can accomplish with an engineered scattering function with distance, $S(x)$, is leveling the distribution over the waveguide length (i.e. waveguide surface area), as discussed next.

Starting with transparent PLA waveguides, we explore two processes to control the scattering in transparent PLA; in the first process we heat PLA in water causing hydrolyzation of the polymer, increasing optical diffusion (Fig. 2) [20]. In the second process NaOH is used as a solvent to PLA to create roughness by etching the PLA surface, which again increases optical diffusion. Both methods result in a biocompatible PLA waveguide with increased optical scattering into the scaffold.

For the second illumination method we introduce an added biocompatible gold film at the end of the PLA waveguide (Fig. 1(c)). The gold acts as a mirror to fold or reflect the light back into the waveguide. Within certain ranges of input power, the reflected light acts to partially compensate for the exponential decay of the original waveguide. The power reflected from the mirror can be modeled as an additional exponential term, except now it decays from the end of the mirror, $L$, back to the source at $x = 0$.

$$P_{mirror}(x) = R A e^{(-\alpha-\sigma)L} e^{(-\alpha-\sigma)(L-x)} \left[ W \right] \quad (5)$$

This leads to a two-term equation for irradiance, $I$, from a gold-mirror waveguide where $R$ is the mirror reflectivity.

$$I(x) = \sigma A \left( e^{(-\alpha-\sigma)x} + R e^{(-\alpha-\sigma)L} e^{(-\alpha-\sigma)(L-x)} \right) \left[ W \cdot m^{-1} \right] \quad (6)$$

The additional exponential term reflecting power from the end of the waveguide acts in the same manner as reflection occurs in a Fabry-Pérot cavity [21, 22]. This levels the illumination over the length of the waveguide (Fig. 1). The gold mirror also effectively sets the power leaving the end of the waveguide to 0, satisfying the second goal of our optimization. The difference between a waveguide's radial power leakage with and without a mirror is summarized in Figure 1. It confirms that with a single reflection the total scattered power is more evenly distributed along the waveguide. This leveling effect continues as the number of reflections in a Fabry-Pérot cavity increases. It is apparent that the leveling of the spatial distribution of the scattered light is proportional to the quality ($Q$) factor of the cavity. For a sufficiently-high $Q$-factor the average scattering function no longer depends on the distance from the source but only on local scattering from within the cavity such as surface roughness, bending, and internal defects.

In our first experiment we treated transparent 3D printer grade filament obtained from BluMat PLA by heating the PLA causing water hydrolyzation of the polymer to increasing optical diffusion while wrapped in damp blotting paper in a microwave oven with varying time intervals. We then attached the treated PLA filaments to a laser light source and measured the scattering and absorption of the PLA filaments (Fig. 2(a)).

To identify the source of the observed increased scattering (Fig. 2), we measured the change in surface roughness of the microwaved PLA via atomic force microscopy (AFM) and measured the change in Raman absorption spectrum of the PLA. The surface roughness was higher in the microwaved samples when compared to the control. The Raman spectrographs also showed an upward shift at the 875 cm$^{-1}$ peak, indicating that both surface roughness and hydrolysis are contributing factors to the increased scattering. In our next experiment we capped the end of the untreated BluMat PLA filament with several micrometers of gold deposited via electron beam deposition. The gold deposition was confined to a cap at the end of the PLA by masking all but the last millimeter of the PLA with Kapton tape. After deposition, we measured the optical scattering from the segments to compare them to PLA segments without gold caps (Fig. 3). Fitting the power decay resutls to eqn (6), we extrated a reflection ratio from the mirror to be approximately $R = 0.7$

In a final third illumination method we build on the cavity-power-leveling effect from the one-dimensional waveguide discussed above, and introduce a ring-resonator mesh as a means to evenly distribute light across a 2-dimensional area (Fig. 4) [24]. A long photon lifetime of a ring resonator results in an equalized spatial illumination over its surface similar to the leaky Fabry-Pérot cavity has an even distribution of light over its surface. The single round trip electric field, $E_1$, becomes a function of the input field [23], $E_i$, via

$$E_1 = \alpha_r e^{j\theta} E_i \qquad (7)$$

where the loss coefficient $\alpha_r$ is for one round trip in the ring and $\exp(jq)$ denotes the change in phase as the wave travels around the circumference of the ring when $q = w2pr/c$. The total electric field, $E_T$, is a sum of the contribution of the electric field from each pass around the ring.

$$E_T = E_i \left[ 1 + \left(\alpha_r e^{j\theta}\right)^1 + \left(\alpha_r e^{j\theta}\right)^2 + \ldots + \left(\alpha_r e^{j\theta}\right)^n \right] \qquad (8)$$

This is a geometric series and for $\alpha < 1$, we can solve for the power at any cross section of the ring.

$$|E_T|^2 = |E_i|^2 \left(\frac{1}{1-\alpha_r}\right)^2 \qquad (9)$$

Assuming losses around the ring exponential decay with distance, we can restate $\alpha_r$ as a function fractional distance along the ring.

$$\alpha_r = e^{(-\alpha-\sigma)2\pi r} \qquad (10)$$

where, as before, $\alpha$ is the absorption coefficient and $\sigma$ is the scattering coefficient. Averaging (10) over lengths greater than $\lambda$ to allow $|e^{j\theta}|^2 = 1$, we can state the irradiance of the ring in terms of material scattering and absorption coefficients.

$$I = \sigma P_i \left(\frac{1}{1-e^{(-\alpha-\sigma)2\pi r}}\right)^2 \qquad (11)$$

As with the leaky Fabry-Pérot cavity, we wish to show that increasing the quality factor decreases spatial variation of the irradiance. To demonstrate this, we restate the electric field offset by some angle $\phi$ around the resonator.

$$E_T(\phi) = e^{(-\alpha-\sigma)\phi r} E_i \left[ 1 + \left(\alpha_r e^{j\theta}\right)^1 + \ldots + \left(\alpha_r e^{j\theta}\right)^n \right]$$

$$= E_i \frac{e^{(-\alpha-\sigma)\phi r}}{1-\alpha_r e^{j\theta}} = E_i \frac{e^{(-\alpha-\sigma)\phi r}}{1-e^{(-\alpha-\sigma)2\pi r} e^{j\theta}} \qquad (12)$$

The irradiance as a function of angle around the ring is a function of input power, and both the absorption and scattering coefficients leading to Eq. (13).

$$I(\phi) = \sigma P_i \left(\frac{e^{(-\alpha-\sigma)\phi r}}{1-e^{(-\alpha-\sigma)2\pi r} e^{j\theta}}\right)^2 \qquad (13)$$

With the aim to minimize the change in irradiance with respect to angle around the ring, towards flattening the irradiance across the ring, we rewrite (13) to become

$$\frac{dI(\phi)}{d\phi} = 2\sigma P_i \left( \frac{e^{-\alpha_T \phi r}}{1 - e^{-\alpha_T 2\pi r} e^{j\theta}} \right)^2 (-\alpha_T) r \qquad (14)$$

With the total loss coefficient $\alpha_T$ defined as $\alpha_T = \alpha + \sigma$. We see that as the total loss coefficient $\alpha_T$ approaches zero, the resulting variation of irradiance over the ring also approaches zero. As light travels around a ring resonator the contribution of scattering from any one interval around the ring becomes minimal and the ring glows evenly, independent of the location of the source. Extending this principle to an array of ring resonators, an entire mesh-like surface can be constructed with an equal scattering distribution provided that coupling losses are minimized. Since the ring $Q$ is a function of both the waveguide width and bending radius, we next explore their influence on equalizing light distribution [24]. A larger bending radius results in a smaller bending loss and vise versa. This principal allows larger and smaller radius and waveguide width rings to be engineered into the mesh to quantitatively vary the illumination at specific locations within the mesh. In this manner if some cells of the scaffold require a higher dose while others require a lower dose, the scaffold itself can be designed to accommodate the cell-dependent variations in optical dose.

To test such equalized light delivery using traveling waves in ring structures, we fabricate a set of four connected ring cavities towards demonstrating cavity-enhanced even spatial light distribution across millimeter areas using by laser cutting PLA film (Fig. 4). The rings are 2.5mm in diameter and formed from a 790μm width waveguide. Light is sourced into the four-ring mesh through a stem of 8.5mm length from a red laser.

The biphasic dose response of cells to optical power in photobiomodulation highlights the necessity of illumination engineering in designing transparent, biocompatible tissue scaffolds. Here we model and demonstrate experimentally three optical methods for leveling the illumination in transparent tissue scaffold structures. While not exhaustive, these three methods: engineering scattering, placement of Fabry-Pérot cavities, and placement of ring resonators; form the underlying principles behind most passive illumination structures. While spatially equalized illumination is only one factor to consider in designing transparent tissue scaffolds, without it, photobiomodulation will be uneven in dose and the tissue scaffold will not reach its full potential. As such engineering optical nanostructures enable next generation devices for with applications in both biomedical treatments and opto-electronics [25-31].

**Figures**

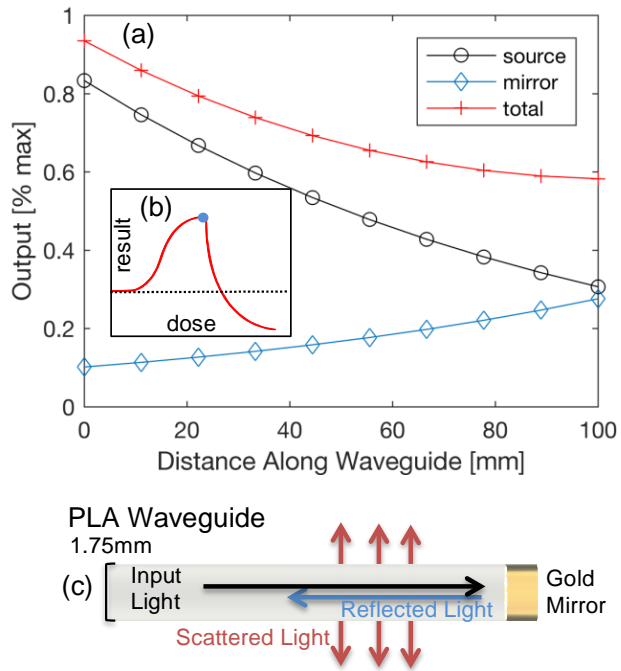

Fig. 1. Modeled irradiance (a) along a 10cm length of PLA without the mirror cap, circles, the contribution of the mirror cap, diamonds, and the total irradiance, crosses, with $a = 10 [m^{-1}]$, $R = 0.9$, shows a leveling of the power distributed over the length of the waveguide in order to reach the optimal biphasic dose response (b), along the length of the fiber, schematic (c).

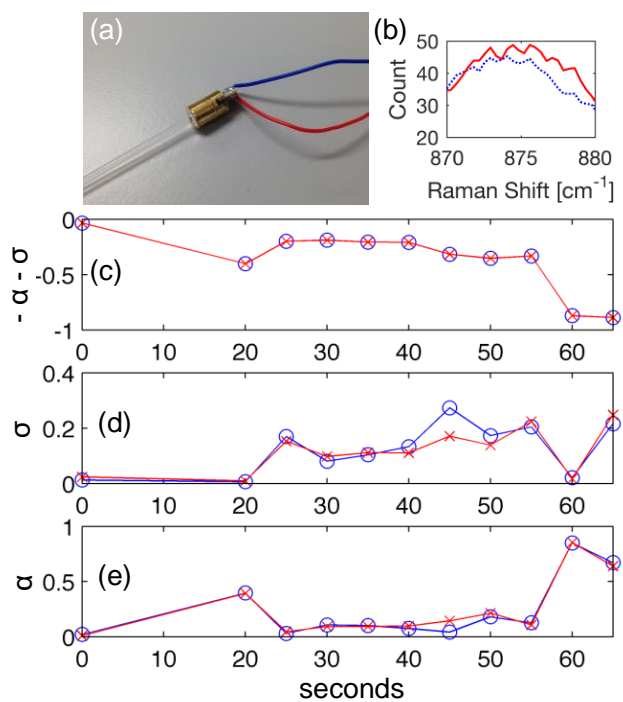

Fig. 2. PLA waveguide absorption ( ) and scattering ) loss test. (a) PLA waveguides were attached to a mini-laser and output power was recorded along with an image from above while the fiber was illuminated. (b) Raman shift of the waveguide prior to treatment (dotted line) and after treatment (solid line) at the 875cm$^{-1}$ peak indicate hydrolysis. PLA waveguides were treated to incur water hydrolyzation of the polymer to increasing optical diffusion through microwave water vaporization. Two measurements of total losses (c) scattering (d) and absorption (e) calculated from curve fitting images and measuring the output power of 11 treated PLA waveguides shows increasing scattering with microwaving time. This light extinction control of these waveguides allows targeting the optimal dosage on the biphasic response curve of cells in biophotomodulation treatments.

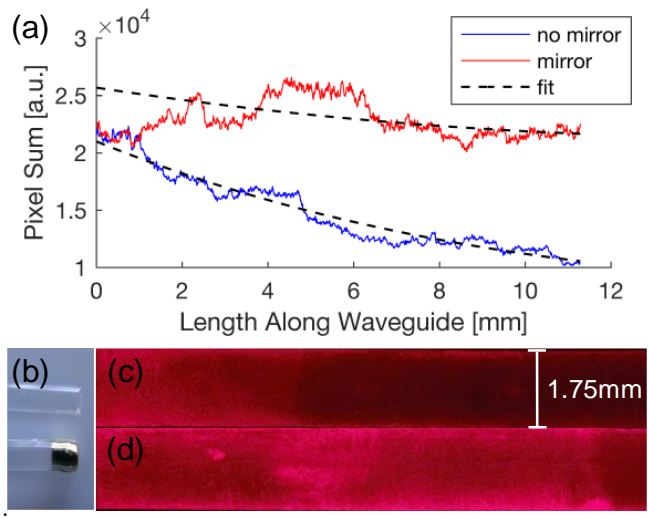

Fig. 3. Light distribution effect from a leaky Fabry-Perot cavity. (a) The cavity is formed on surface-treated PLA fibers where one end is capped with a mirror (100 nm of gold). The mirror folds light back into the fiber, thus distributing the energy more evenly along the fiber. (b) Gold mirror cap. (c) Image of the illuminated PLA fiber with out gold mirror. (d) Image of the illuminated PLA fiber with gold mirror.

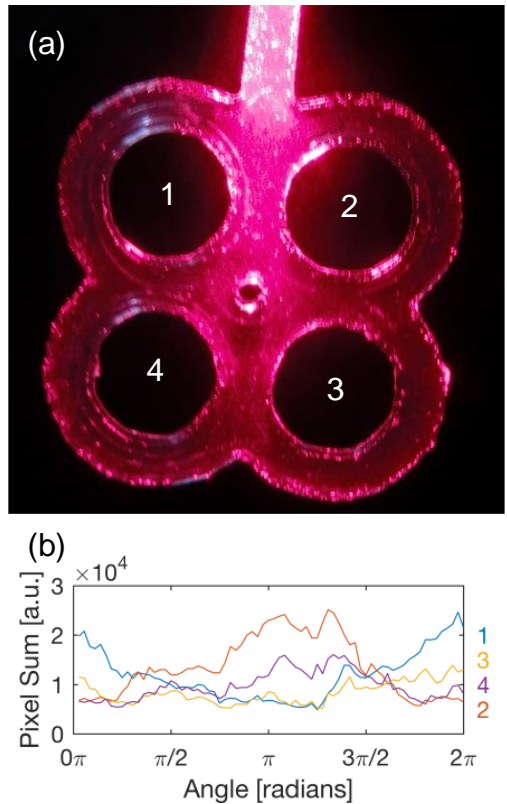

Fig. 4. A four-ring mesh fabricated from transparent 180μm thick PLA film shows leveling of the distribution of power over the surface of the structure (a). The intensity plotted with angle where the angle 0 is the left most point in each ring, shows an average 4.9dB variance in irradiance around each ring (b), limited by the rough edges of the laser cut PLA.